\documentclass[runningheads]{llncs}
\usepackage{graphicx}
\usepackage{amsmath}
\usepackage{booktabs}
\usepackage{algorithm}
\usepackage{algpseudocode}
\usepackage{todonotes}
\usepackage{array}
\usepackage{xcolor}
\usepackage{ulem}
\usepackage{hyperref}
\newcolumntype{L}[1]{>{\raggedright\let\newline\\\arraybackslash\hspace{0pt}}m{#1}}
\newcolumntype{C}[1]{>{\centering\let\newline\\\arraybackslash\hspace{0pt}}m{#1}}
\newcolumntype{R}[1]{>{\raggedleft\let\newline\\\arraybackslash\hspace{0pt}}m{#1}}
\begin{document}
\newif\ifanonymize
\anonymizefalse

%
%

\ifanonymize
    \title{Content-Aware Differential Privacy with Conditional Invertible Neural Networks}
    \author{Anonymous\inst{1}}
    \institute{Anonymous Insitution}
\else
    \title{Content-Aware Differential Privacy with Conditional Invertible Neural Networks
    }
    \author{Malte Tölle\inst{1,3,4}
    \and
    Ullrich Köthe\inst{2,3}
    \and
    Florian André\inst{1,3,4}
    \and
    Benjamin Meder\inst{1,3,4}
    \and
    Sandy Engelhardt\inst{1,3,4}
    }
    \authorrunning{M. Tölle et al.}
    \institute{
    Department of Internal Medicine III, Heidelberg University Hospital
    \and
    Visual Learning Lab, Ruprecht-Karls University Heidelberg\\
    \and
    Informatics for Life Institute, Ruprecht-Karls University Heidelberg \\
    \and
    DZHK (German Centre for Cardiovascular Research), partner site Heidelberg/Mannheim \\
    \email{malte.toelle@med.uni-heidelberg.de}}
\fi

\titlerunning{Content-Aware Differential Privacy}
%

%
\maketitle              
\begin{abstract}
Differential privacy (DP) has arisen as the gold standard in protecting an individual's privacy in datasets by adding calibrated noise to each data sample. 
While the application to categorical data is straightforward, its usability in the context of images has been limited. 
Contrary to categorical data the meaning of an image is inherent in the spatial correlation of neighboring pixels making the simple application of noise infeasible.
Invertible Neural Networks (INN) have shown excellent generative performance while still providing the ability to quantify the exact likelihood. 
Their principle is based on transforming a complicated distribution into a simple one e.g.\ an image into a spherical Gaussian.
We hypothesize that adding noise to the latent space of an INN can enable differentially private image modification.
Manipulation of the latent space leads to a modified image while preserving important details.
Further, by conditioning the INN on meta-data provided with the dataset we aim at leaving dimensions important for downstream tasks like classification untouched while altering other parts that potentially contain identifying information.
We term our method \textit{content-aware differential privacy} (CADP).
We conduct experiments on publicly available benchmarking datasets as well as dedicated medical ones.
In addition, we show the generalizability of our method to categorical data.
The source code is publicly available at \url{github.com/Cardio-AI/CADP}.

\end{abstract}

\section{Introduction}

The predictive performances of algorithms especially neural networks are heavily dependent on the amount of data they are trained with.
In contrast, privacy regulations aiming at hiding individual sensitive information hinder the application of machine learning tools on heterogeneous multi-center data. 
Since it is not our objective to argue about the benefits of these privacy regulations,
we strive to find methods that allow publishing of sensitive data simultaneously to maintaining individual's privacy.
While such methods are trivial to implement for 
categorical data (e.g.\ a data base with entries for sex, age, gender, etc.) complex data such as images pose a difficult objective.
Contrary to categorical data images obtain their meaning by the spatial relationship of individual pixels.
Perturbing pixels by adding random noise would not hinder a human or a machine observer from re-identifying the image's content; recognizing people by their face being the most obvious example.
Older techniques rely on blurring or pixelation of people's faces, e.g.\ Google Street View \cite{frome2009google}.

\begin{figure}[H]
    \centering
    \includegraphics[width=\textwidth]{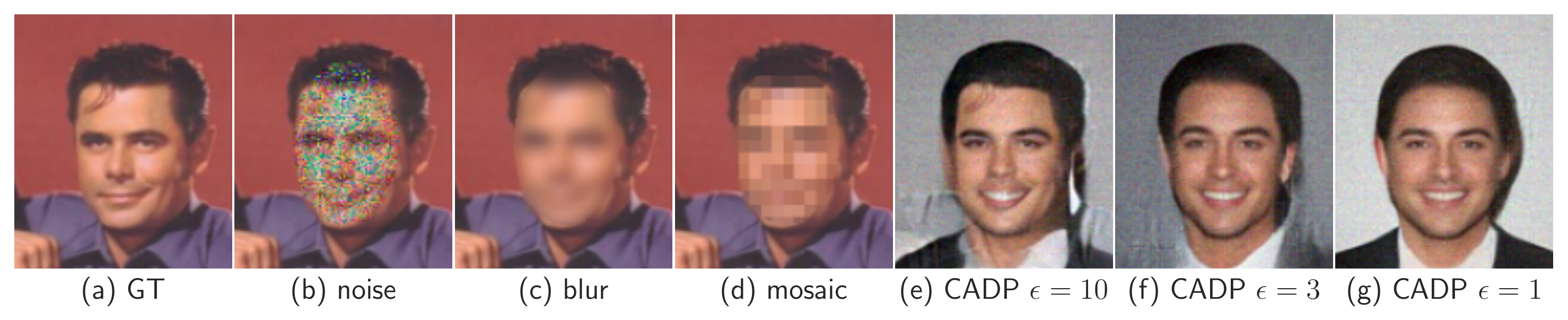}
    \caption{Example of face anonymization with Differential Privacy \cite{liu2015celeba}. 
    Compared to conventional approaches based on noise (a), blur (b), and mosaic (d) our content-aware approach (e)-(g) changes the identity of the image. 
    For $\epsilon=10$ (e) one can still see strong similarities between reconstruction and ground truth as e.g.\ the lock of hair on the forehead. 
    For small $\epsilon$ the similarity decreases as desired to disable re-identification.
    However, if the subsequent task was to classify the eye color, this would still be possible with the CADP results from (e)-(g), since we can condition the transformation and therefore leave important aspects unaltered.
    }
    \label{fig:face_example}
    \vspace{1cm}
    \centering
    \includegraphics[width=0.8\textwidth]{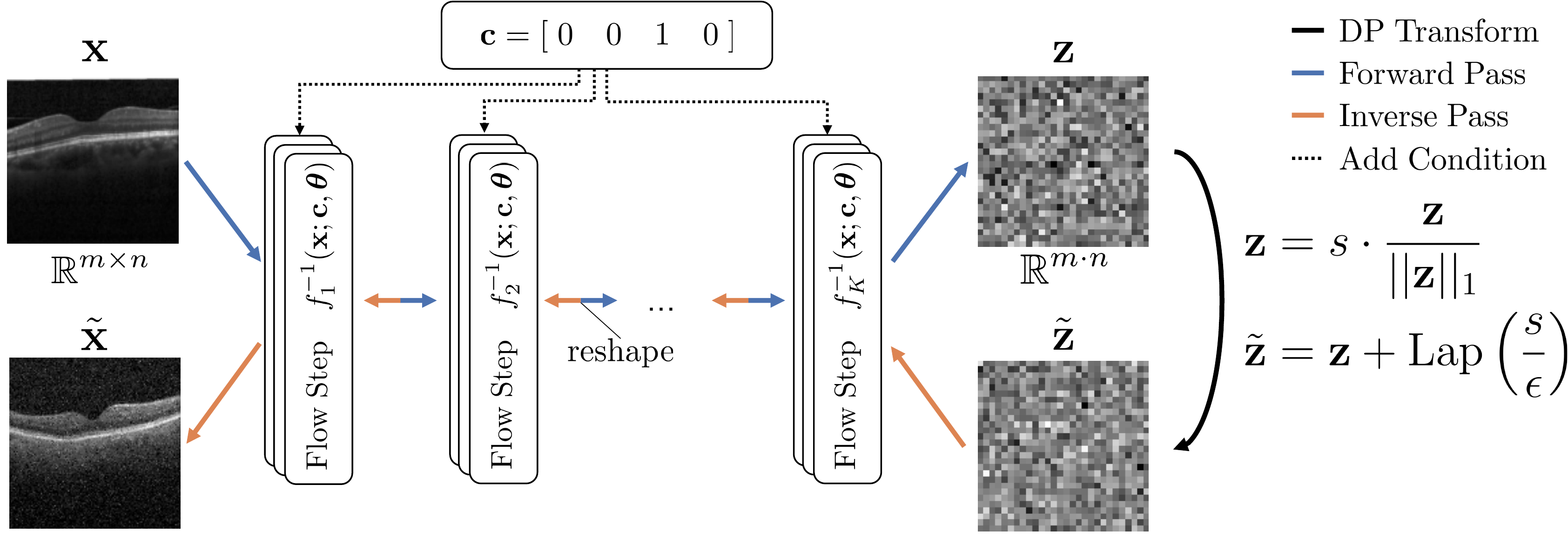}
    \caption{Content-aware differential privacy (CADP) pipeline. After training the INN to convergence we feed each sample $\mathbf{x}$ with the corresponding condition $\mathbf{c}(\mathbf{y})$ to obtain our latent representation $\mathbf{z}$. After clipping its $L_1$-norm to the desired sensitivity $s$, Laplacian distributed noise $\mathrm{Lap}(0,s/\epsilon)$ is added to obtain $\epsilon$-DP. The perturbed $\tilde{\mathbf{z}}$ is fed in reverse to obtain the differentially private image $\tilde{\mathbf{x}}$.}
    \label{fig:idea_cadp}
\end{figure}

Training of machine learning models with such samples would tremendously decrease their predictive performance because a great deal of features are lost in the process which the model never sees (see Fig.\ \ref{fig:face_example}).
This is of utmost importance in the medical domain as we must ensure the model learns on valid features for detecting pathologies.

We hypothesize that the tools of machine learning namely neural networks based on Normalizing Flows (NF) known as Invertible Neural Networks (INN) may be used to address the privacy issue when dealing with images and medical ones in particular \cite{ardizzone2018inn}.
Our contribution is three-fold:
\begin{itemize}
    \item First, we provide mathematically grounded evidence that INNs provide a valuable tool to obtain $\epsilon$-differentially private images that exhibit all features of natural images (e.g.\ sharpness or authenticity).
    $\epsilon$ quantifies the probability of data leakage, the lower $\epsilon$ the more privacy is guaranteed.
    \item Second, by conditioning our network on meta-data provided in conjunction with the dataset (e.g.\ pathologies) the INN is able to automatically extract dimensions most likely corresponding to classifying those meta variables.
    We assume these features merit attention for downstream tasks and, thus, should be modified as little as possible self-evident within the bounds of desired privacy. 
    We term this method \textit{Content-Aware DP} (CADP). 
    \item Third, we show the generalizability of our method not just to images but also to categorical data making it a universal tool for obtaining differentially private data.
\end{itemize}

We focus on the task of protecting images in particular, or data in general in any context, detached from their intended usage.
    


\section{Related Work}

\paragraph{Differentially Private Invertible Neural Networks.} 
In general each learning based algorithm can be trained in a privacy preserving fashion by using differentially private stochastic gradient descent (DP-SGD) \cite{abadi2016dpsgd}.
DP-SGD achieves differentially private model training by clipping the per-sample gradient and adding calibrated Gaussian noise proportional to the desired level of privacy.
Therefore, DP-SGD tweaks the model parameters instead of the input to obtain privacy by e.g.\ ensuring no inputs might be reconstructed from the model parameters \cite{usynin2021attacks}.

One can distinguish between input-, output-, and algorithm-perturbation to achieve DP.
When the output of the algorithm or the algorithm itself is perturbed as e.g.\ in DP-SGD the analysis is performed on the non-private data, where one has to be concerned about the composition property ($\epsilon$ degrades over multiple analyses of the dataset).
Further, since one cannot release the data the possibilities for analysis are limited.
We circumvent above mentioned limitations by performing input-perturbation and use the robustness of DP against post-processing (any further processing of differential private data retains privacy guarantees).

Obviously, INNs can be trained with DP-SGD as well \cite{waites21dpnf}. 
However, after training one can only use the INN in a generative manner by sampling the latent space $\mathbf{z} \sim \mathcal{N}(\mathbf{0};\mathbf{I})$ and obtain data samples that have no relation to in reality occuring data samples and are therefore artificial.
Thus, it does not allow for perturbation of the real data samples intended to be published or used for model training.
Even worse, using artificial data is also not completely secure against attacks \cite{bellovin2018failgan} and may even lead to wrong pathologies in generated images \cite{Bhadra2021OnHI,Laves2020MCDIP}.

\paragraph{Differential Privacy for Images.} 
The most prominent application in the literature about differentially private images deals with faces, as this is the most vivid example.
Older approaches rely on pixeling, blurring, obfuscation, or inpainting \cite{Fan2018ImagePW}, but this has been proven as ineffective against deep learning based recognizers \cite{mcpherson2016defeating,oh2016personrecognition}.
Another promising path is the generation of fully artificial data with e.g.\ Generative Adversarial Networks (GAN) with the known drawbacks mentioned above \cite{bissoto2018ganskinlesion,schuette2021gansynthetic,waites21dpnf,yoon2018pategan}.
Ziller et al. claimed to having applied DP to medical images. \cite{ziller2021dpsgd}.
However, their approach also only involves trainng a conventional CNN on medical images with DP-SGD.
We take a different path and \textit{alter the content of the input image} in a private manner as we want to preserve as much information as possible and only alter dimensions that are not identification related.
To the best of our knowledge DP has never been applied \textit{directly to the content of medical images} before.


\section{Methods}

\subsection{(Conditional) Invertible Neural Networks}

INNs deal with the approximation of a complex, unobservable distribution $p(\mathbf{x})$ by a simpler tractable prior $q(\mathbf{z})$, usually a spherical multivariate Gaussian. 
Let $\mathcal{X} = \left\{ \mathbf{x}^{(1)}, ..., \mathbf{x}^{(n)} \right\}$ be $n$ observed i.i.d. samples from $p(\mathbf{x})$.
The objective is to approximate $p(\mathbf{x})$ via a model $f_{\boldsymbol{\theta}}$ consisting of a series of $K$ bijective functions $f_{\boldsymbol{\theta}} = f_1 \odot ... \odot f_K$ parameterized fully by $\boldsymbol{\theta}$ transforming $q(\mathbf{z})=\mathcal{N}(\mathbf{0};\mathbf{I})$ into $p(\mathbf{x})$ and vice versa ($f_{\boldsymbol{\theta}}(\mathbf{x})=\mathbf{z}\longleftrightarrow f_{\boldsymbol{\theta}}^{-1}(\mathbf{z})=\mathbf{x}$).

Such a model can efficiently be used in a generative manner to sample $\mathbf{x} \sim p$ by first sampling $\mathbf{z} \sim \mathcal{N}(\mathbf{0};\mathbf{I})$ and subsequently transforming the sample as $\mathbf{x} = f_{\boldsymbol{\theta}}(\mathbf{z})$.

    

\noindent Since $f_{\boldsymbol{\theta}}$ exhibits invertibility, exact likelihood evaluation becomes tractable by utilizing the change of variables formula \cite{dinh2015nice,dinh2017realnvp}.

\begin{equation}
    \begin{split}
    \log p(\mathbf{x}) &= \log q \left( f_{\boldsymbol{\theta}}^{-1} (\mathbf{z}) \right) + \log \left| \det \left( \frac{\partial f_{\boldsymbol{\theta}}^{-1}(\mathbf{z})}{\partial \mathbf{x}} \right) \right| 
    \end{split}
\end{equation}

\noindent An isotropic Gaussian is usually chosen as prior. 
Since its covariance matrix is diagonal, components are independent. 
With INNs sharp image details can be obtained, while simultaneously allowing to modify independent components of the image in latent space \cite{kingma18glow}. 

We build on the foundations laid by Ardizzone et al., who incorporated conditions  by e.g.\ concatenation of class labels to the input \cite{ardizzone19cinn}.
This enables the INN to implicitly learn the meta-data dependent distribution in latent space.
In the reverse pass we provide the label we would like to obtain, e.g.\ a pathology, and the INN generates an altered version of the original image that still exhibits the desired pathology ($f_{\boldsymbol{\theta}}(\mathbf{x},\mathbf{c})=\mathbf{z}\longleftrightarrow f_{\boldsymbol{\theta}}^{-1}(\mathbf{z},\mathbf{c})=\mathbf{x}$).




\subsection{Content-Aware Differential Privacy}
\label{sec:dp}

Being termed the gold standard in obscuring data sample sensitive information, DP provides a mathematically grounded, quantifiable measure of leaked information while simultaneously being applicable in a simple manner 
\cite{ziller2021deepee}.
From a high-level perspective it guarantees that changing one value in the database ($\mathcal{X}$ and $\mathcal{X}^{\prime}$) will have only a small effect on the model prediction \cite{dwork2014foundationDP}.

    \begin{equation}
        Pr \left[ \mathcal{M}(\mathcal{X}) \in \mathcal{S} \right] \leq \exp(\epsilon) Pr \left[ \mathcal{M}(\mathcal{X}^{\prime}) \in \mathcal{S} \right] \; ,
    \end{equation}

\noindent where $\mathcal{M}$ denotes a randomized mechanism and $\mathcal{S}$ all sets of outputs.
The closer the two probabilities are, the less information is leaked (small $\epsilon$). 
DP is usually obtained by perturbing data with calibrated noise proportional to the function's $f$ ($L_1$-norm) sensitivity on dataset $\mathcal{X}$, which is the maximum change in the function's value by changing one data point.
To achieve pure $\epsilon$-DP the Laplace mechanism is commonly used.
\vspace{-.6cm}
\begin{multicols}{2}
    \begin{equation}
        s = \max_{\mathcal{X},\mathcal{X}^{\prime}} || f(\mathcal{X}) - f(\mathcal{X}^{\prime}) ||_1 \;,
    \end{equation}\break
    \begin{equation}
        \mathcal{M}(\mathcal{X}) = f(\mathcal{X}) + \mathrm{Lap} \left( \frac{s}{\epsilon} \right) \; .
    \end{equation}
\end{multicols}

noindent After training an INN to convergence i.e.\ $f_{\boldsymbol{\theta}}(\mathcal{X},\mathcal{C}) \sim \mathcal{N}(\mathbf{0},\mathbf{I})$, each image and label $(\mathbf{x}_i,\mathbf{y}_i)\in\mathcal{X}$ with corresponding condition $\mathbf{c}_i(\mathbf{y}_i)$ is forwarded through the network (see Fig. \ref{fig:idea_cadp}).
The resulting latent space $f_{\boldsymbol{\theta}}(\mathbf{x}_i,\mathbf{c}_i(\mathbf{y}_i)) = \mathbf{z}_i$ is modified in a differentially private manner by sampling from a Laplace distribution with standard deviation determined by the sensitivity $s$ and the desired $\epsilon$.
We clip our sensitivity by dividing each $\mathbf{z}_i$ by its $L_1$-norm (Alg.\ \ref{alg:cadp}) \cite{abadi2016dpsgd}.
Since $\mathcal{Z}$ is learned to be an isotropic Gaussian each component is independent and can, thus, be modified individually.
INNs can trivially be expanded to be trained on categorical data as well, making our method a general technique for applying DP on data.

\begin{theorem}[$\epsilon$-Content-Aware-DP Mechanism]
    For an image $\mathbf{x} \in \mathcal{X}$ there exists a mechanism $\mathcal{M}_{\mathrm{CA}}$ that maps $\mathbf{x}$ to its differentially private counterpart $\tilde{\mathbf{x}} \in \mathcal{X}$. We say $\mathcal{M}_{\mathrm{CA}}$ satisfies $\epsilon$-DP, if and only if for all $\mathbf{x},\mathbf{x}^{\prime} \in \mathcal{X}$
    \begin{equation}
        \mathcal{M}_{\mathrm{CA}} = f_{\boldsymbol{\theta}}^{-1} \left[ f_{\boldsymbol{\theta}}(\mathbf{x}) + (l_1, ... , l_k) \right] = f_{\boldsymbol{\theta}}^{-1} \left[\mathbf{z} + (l_1, ... , l_k) \right] = f_{\boldsymbol{\theta}}^{-1} \left[\tilde{\mathbf{z}} \right] \;,
    \end{equation}
    where $f_{\boldsymbol{\theta}}$ denotes a function that maps $\mathbf{x}$ to a latent vector $\mathbf{z} \in \mathcal{Z}$ and by reverse pass $f_{\boldsymbol{\theta}}^{-1}$ maps $\mathbf{z}$ to $\mathbf{x}$. $\tilde{\mathbf{z}} = \mathbf{z} + (l_1,...,l_k)$ denotes the $\epsilon$-DP perturbed version of $\mathbf{z}$ with $l_i$ i.i.d.\ random variables drawn from $\mathrm{Lap}\left( s / \epsilon\right)$. 
\end{theorem}

    

\begin{proof}
    Let $\mathbf{x} \in \mathcal{R}^{|\mathcal{X}|}$ and $\mathbf{x}^{\prime} \in \mathcal{R}^{|\mathcal{X}|}$ be such that $|| \mathbf{x} - \mathbf{x}^{\prime} ||_1 \leq 1$, and $g(\mathbf{x}) = f_{\boldsymbol{\theta}}^{-1}\left( f_{\boldsymbol{\theta}} (\mathbf{x}) \right)$ be some function $g: \mathcal{R}^{|\mathcal{X}|} \rightarrow \mathcal{R}^{|\mathcal{Z}|} \rightarrow \mathcal{R}^{|\mathcal{X}|}$. 
    We only consider functions that are volume preserving meaning their Jacobian determinant is equal to one $\left( \left| \det\left( \partial f_{\boldsymbol{\theta}}(\mathbf{x}) /\partial \mathbf{z} \right) \right| = 1 \right)$.
    Let $p_{\mathbf{x}}$ denote the probability density function of $\mathcal{M}_{\mathrm{CA}}(\mathbf{x},g,\epsilon)$, and $p_{\mathbf{x}^{\prime}}$ of $\mathcal{M}_{\mathrm{CA}}(\mathbf{x}^{\prime},g,\epsilon)$. 
    We assume the distance between points is similar in $\mathcal{X}$ and $\mathcal{Z}$ as shown by \cite{kingma18glow}.
    We compare the two at some arbitrary point $\mathbf{t} \in \mathcal{R}^{|\mathcal{Z}|}$
    \begin{equation}
        \begin{aligned}
            \frac{p_{\mathbf{x}}(\mathbf{t})}{p_{\mathbf{x}^{\prime}}(\mathbf{t})} &= \prod_{i=1}^{k} \left( \frac{ \exp \left( - \frac{\epsilon}{s} | g(\mathbf{x}) - f_{\boldsymbol{\theta}}^{-1} (\mathbf{t}) | \right) } { \exp \left( - \frac{\epsilon}{s} | g(\mathbf{x}^{\prime}) - f_{\boldsymbol{\theta}}^{-1} (\mathbf{t}) | \right) } \right)
            = \prod_{i=1}^{k} \left( \frac{ \exp \left( - \frac{\epsilon}{s} | f_{\boldsymbol{\theta}}^{-1} \left( f_{\boldsymbol{\theta}} (\mathbf{x}) - \mathbf{t} \right) | \right) } { \exp \left( - \frac{\epsilon}{s} | f_{\boldsymbol{\theta}}^{-1} \left( f_{\boldsymbol{\theta}} (\mathbf{x}^{\prime}) - \mathbf{t} \right) | \right) } \right)\\
            &= \prod_{i=1}^{k} \left( \exp - \frac{\epsilon}{s} |f_{\boldsymbol{\theta}}^{-1}\left( \mathbf{z}_{\mathbf{x}} - \mathbf{t} \right) - f_{\boldsymbol{\theta}}^{-1}\left( \mathbf{z}_{\mathbf{x}^{\prime}} - \mathbf{t} \right)| \right)\\
            &= \prod_{i=1}^{k} \left( \exp - \frac{\epsilon}{s} |f_{\boldsymbol{\theta}}^{-1}\left( \mathbf{z}_{\mathbf{x}} -  \mathbf{z}_{\mathbf{x}^{\prime}} \right)| \right)\\
            &\leq \prod_{i=1}^{k} \exp \left( - \frac{\epsilon | \mathbf{z}_{\mathbf{x}} - \mathbf{z}_{\mathbf{x}^{\prime}} |}{s} \right)
            = \exp \left( \frac{\epsilon || \mathbf{z}_{\mathbf{x}} - \mathbf{z}_{\mathbf{x}^{\prime}} ||_{1}}{s} \right)\\
            &\leq \exp (\epsilon) \;,
        \end{aligned}
    \end{equation}
    where the first inequality follows from the triangle inequality, and the last follows from the definition of sensitivity and $||\mathbf{x}-\mathbf{x}^{\prime}||_1 \leq 1$. $\frac{p_{\mathbf{x}}(\mathbf{t})}{p_{\mathbf{x}^{\prime}}(\mathbf{t})} \geq \exp (- \epsilon)$ follows by symmetry. 
\end{proof}

\section{Experiments}

We apply our approach for content-aware differential privacy to several publicly available datasets to showcase its generalizability. 
In each case we first train the INN on the training partition and subsequently train a classifier on the differentially private data.
Note that our goal is not to reach as high as possible predictive performance but to close the gap between original and differentially private training.
To exemplify the principle of content-aware DP we use the MNIST dataset, since the effect of transformations in latent space is obvious \cite{lecun2010mnist}. 
Next, we use two dedicated medical datasets, the first being a collection of retinal optical coherence tomography (OCT) scans with four classes (choroidal neovascularization (CNV), diabetic macular edema (DME), drusen, and healthy) \cite{kermanydata} and the second being a series of chest x-ray scans with healthy and pneumonic patients \cite{kermanydata}, which contain more complicated and indistinct transformations.

Since most works in adding privacy to images deal with the prototype example of identifiability of faces, we also apply our approach to the CelebA Faces dataset (see Fig.\ \ref{fig:face_example}) \cite{liu2015celeba}.
After having investigated our method on image data, we expand it to categorical data i.e.\ diabetes dataset from \texttt{scikit-learn} \cite{scikit-learn}.

For each dataset we train a separete INN with convolutional subnetworks, with depth (number of downsampling operations) dependent on the image resolution. 
We chose $d=2$ for MNIST ($28 \times 28$), $d=4$ for OCT and chest x-ray ($128 \times 128$), and $d=6$ for CelebA ($3 \times 128 \times 128$). 
As coupling block we use the volume preserving GIN (general incompressible-flow) \cite{Sorrenson2020Disentanglement} for MNIST and diabetes data, and Glow (generative flow) \cite{kingma18glow} for the other, more complicated datasets. 
After having trained an INN to convergence we train a classifier with 

{
    \centering
\begin{minipage}{.52\linewidth}
    \begin{figure}[H]
    \includegraphics[width=0.95\textwidth]{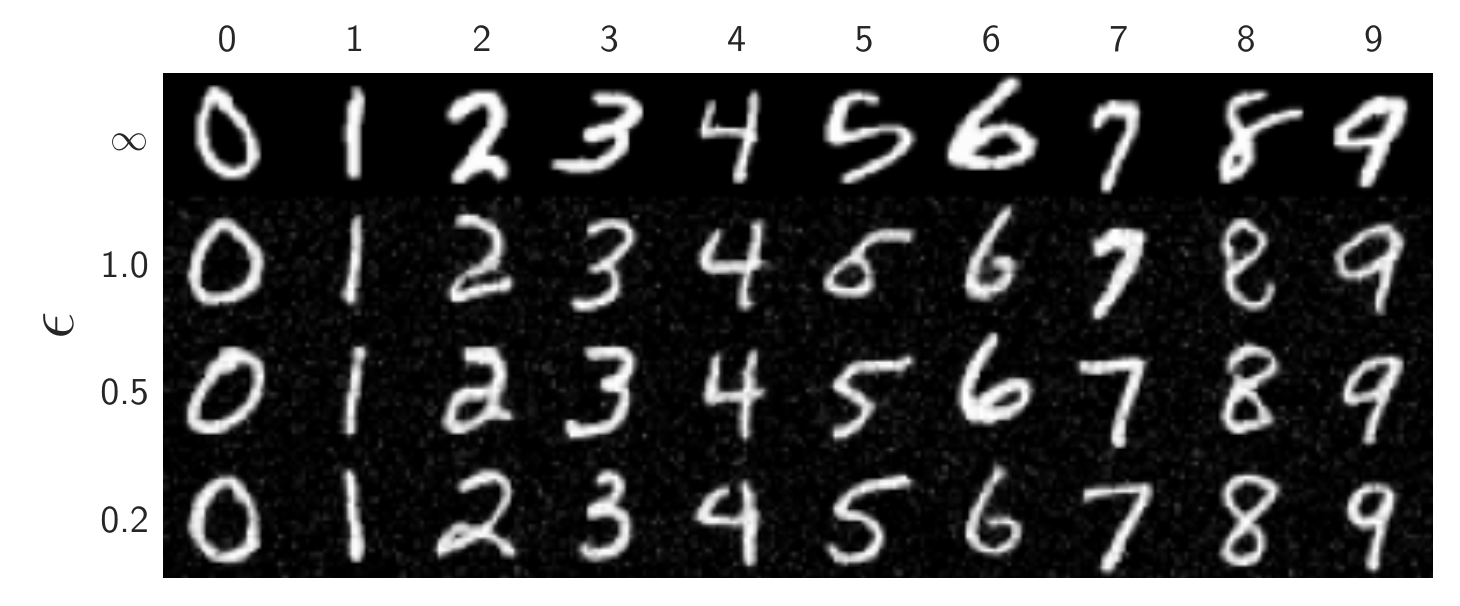}
    \vspace{-.2cm}
    \caption{Differentially private reconstruction of MNIST with different $\epsilon$ and $s=\epsilon/2$.}
    \label{fig:mnist_laplace_dp}
    \end{figure}
    \vspace{-1.1cm}
    \begin{figure}[H]
    \includegraphics[width=\textwidth]{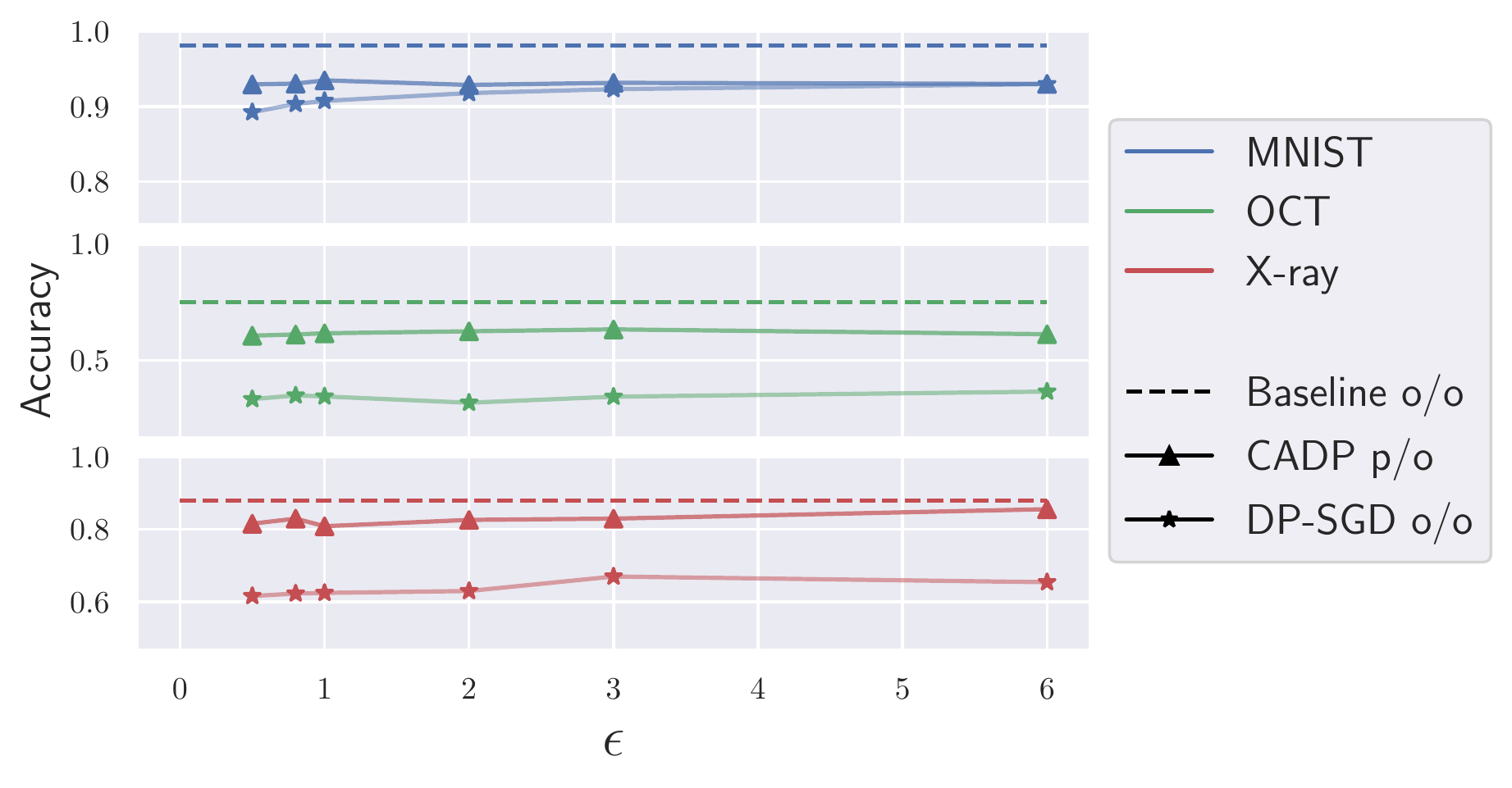}
    \vspace{-.8cm}
    \caption{Accuracy of classifier on different datasets with different $\epsilon$ and $s=\mathrm{min}(\epsilon/2,4)$. 
    Further, we trained the same model with DP-SGD \cite{abadi2016dpsgd}.
    Training/testing is performed on either original (o) or CADP altered (p) data.
    }
    \label{fig:classifier}
    \end{figure}
\end{minipage}
\hspace{.5cm}
\begin{minipage}{.42\linewidth}
    \begin{algorithm}[H]
        \caption{CADP}\label{alg:cadp}
        \begin{algorithmic}
            \Require Samples from training set $\mathcal{X} = \{(\mathbf{x}_1, \mathbf{y}_1),...,(\mathbf{x}_N,\mathbf{y}_N)\}$ with corresponding conditions $\mathcal{C}=\{\mathbf{c}_1(\mathbf{y}_1), ..., \mathbf{c}_N(\mathbf{y}_N) \}$, INN $f_{\boldsymbol{\theta}}$ trained to convergence s.t.\ $f_{\boldsymbol{\theta}}(\mathcal{X}) = \mathcal{Z} \sim \mathcal{N}(\mathbf{0},\mathbf{I})$, sensitivity $s$, epsilon $\epsilon$
            \For{$(\mathbf{x}_i, \mathbf{y}_i) \in \mathcal{X}$ and $\mathbf{c}_i(\mathbf{y}_i) \in \mathcal{C}$}
                \State \textbf{Forward pass}
                \State $\mathbf{z}_i \gets f_{\boldsymbol{\theta}}(\mathbf{x}_i, \mathbf{c}_i(\mathbf{y}_i))$
                \State \textbf{Clip norm of $\mathbf{z}_i$}
                \State $\mathbf{z}_i \gets s \cdot \frac{\mathbf{z}_i}{|| \mathbf{z}_i ||_1}$
                \State \textbf{Add calibrated noise}
                \State $\tilde{\mathbf{z}}_i \gets \mathbf{z}_i + \mathrm{Lap}\left( \frac{s}{\epsilon} \right)$
                \State \textbf{Reverse Pass}
                \State $\tilde{\mathbf{x}}_i \gets f_{\boldsymbol{\theta}}^{-1} (\tilde{\mathbf{x}}_i, \mathbf{c}_i(\mathbf{y}_i))$
            \EndFor\\
            \hspace{-.3cm}\textbf{Output:}\\ $\tilde{\mathcal{X}} = \{(\tilde{\mathbf{x}}_1, \mathbf{y}_1),...,(\tilde{\mathbf{x}}_N, \mathbf{y}_N) \}$ 
        \end{algorithmic}
    \end{algorithm}
\end{minipage}%
\vspace{1cm}
}

\noindent convolutional blocks 
and two linear layers on the differentially private data.
Testing is performed on original data to investigate the amount of true features the model learns.
We believe that the performance of the classifier acts as an implicit benchmark to make sure the INN not only reconstructs conditional noise.
It is common practice for all works dealing with DP algorithms to be compared to the non-private benchmark. 
The goal must be to close the still existing gap to incentivize differentially private training by eliminating all its shortcomings.
For comparison we also train the same classifier with DP-SGD, the current gold standard \cite{abadi2016dpsgd}.
All experiments were performed on a NVIDIA Titan RTX.

\section{Results}
The results are presented in a two-fold manner.
We first show the differentially private adjusted images per class for each dataset with different levels of $\epsilon$.
Second, we show the reached accuracy of the classifier on the original, not-CADP altered test data chunk when trained on the original, on the CADP altered dataset, or with DP-SGD. 
\paragraph{MNIST.}
Even for small $\epsilon$ our approach generates visually appealing results that are indistinguishable from real digits but exhibit a large difference from the original (see Fig.\ \ref{fig:mnist_laplace_dp}).
Attributes being altered are line thickness (e.g.\ 6), slant (e.g.\ 1), and even style (e.g.\ 2).
For $\epsilon=0.2$ a classifier trained on CADP-altered data outperforms the commonly accepted DP-SGD, CADP reaches 92.94\% accuracy while DP-SGD only results in 89.24\% (c.f.\ Fig.\ \ref{fig:classifier}).
The gap closes for larger $\epsilon$.

\begin{figure}[H]
    \centering
    \includegraphics[width=\textwidth]{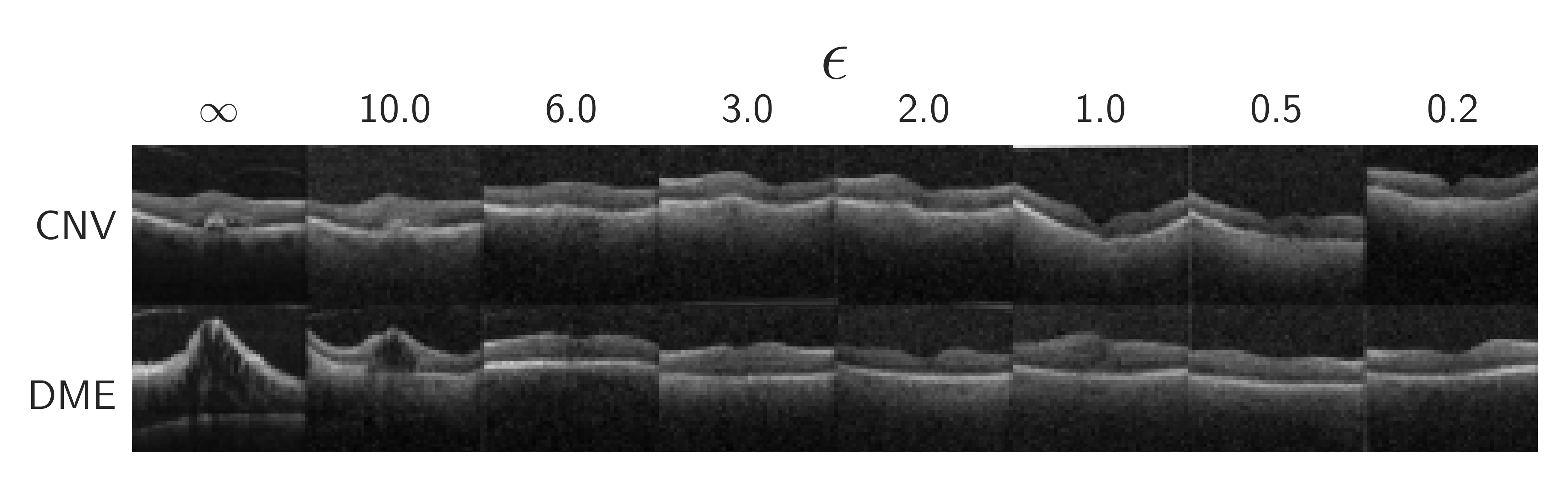}
    \caption{Content-aware differentially private images from OCT dataset with different $\epsilon$ for classes \textit{CNV} and \textit{DME} \cite{kermanydata}. 
    The sensitivity is set to $\mathrm{min\,}(\epsilon/2, 4)$.
    For high $\epsilon$ (e.g. 10) the reconstructed retinal OCT still share similarities as in Fig. \ref{fig:face_example}.
    For smaller $\epsilon$ qualitatively the images look different from their original counterpart.
    However, the classifier (Fig. \ref{fig:classifier}) still performs well acting as an implicit control of the preserved features.
    }
    \label{fig:oct_laplace_dp}
    \vspace{.6cm}
    \centering
    \includegraphics[width=\textwidth]{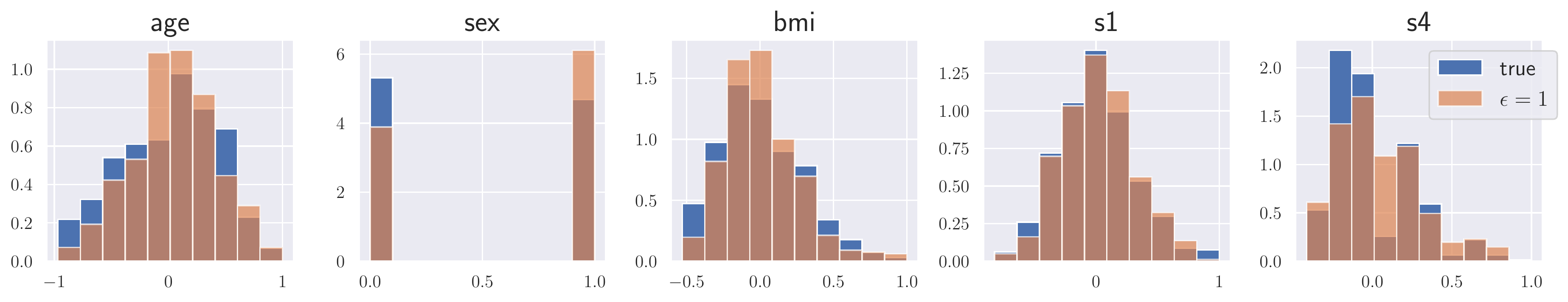}
    \caption{Content-aware differentially private data from diabetes dataset from \texttt{scikit-learn}  with $\epsilon=1$ and sensitivity $s=1$ \cite{scikit-learn}. With conditions the INN is able to reconstruct the approximate distributions even if binary distributed.}
    \label{fig:categorical_dp}
\end{figure}

\paragraph{Retinal OCT and Chest X-ray.}
In retinal OCTs the perturbations are rather subtle and difficult to interpret for a human observer or a non-expert.
Identification related attributes like retinal detachments in specific places are (re-)moved impeding de-identification (see Fig.\ \ref{fig:oct_laplace_dp}).
The CADP-altered images images exhibit transformations resulting in large dissimilarites to their original counterpart.
However, CADP induces a smaller privacy-utility tradeoff since the performance of the classifier trained on CADP altered data is close to the one trained on original data (Fig.\ \ref{fig:classifier}).
The classifier trained on data altered by our method outperforms the one trained with DP-SGD by 23.63\% on average across all $\epsilon$ on the OCT test dataset and by 16.52\% on the chest X-ray test dataset.
We attribute this to the content-awareness of our method, which leaves dimensions corresponding to conditions, i.e.\ pathologies, unaltered.
This is desirable in settings, where one trains a model on private data of another location, e.g.\ a hospital, and applies it to its own in-house samples.

\paragraph{Categorical Data.}
INNs can also generate differentially private categorical data as can be seen in Fig.\ \ref{fig:categorical_dp} for the diabetes dataset from \texttt{scikit-learn} \cite{scikit-learn}.
The data distributions are kept similar but are still altered equipping each data sample with plausible deniability.
To obtain the binary feature of sex, we condition the INN on this feature; the others are learned in an unsupervised fashion.

\section{Discussion and Conclusion}

We introduced a new method to achieve differentially private images based on invertible neural networks, which we term CADP (content-aware differential privacy).
We applied the method to medical images and ensured the identity i.e.\ pathology of the patient is not changed by conditioning the INN on the class labels.
We could show that in three experiments on diverse medical data (images of digits, OCT, and X-ray scans), the subsequent classifiers outperformed conventional approaches by a margin when fed with CADP-generated data.
By this we reduce the risk for false diagnosis and increase the safety of patients against wrong diagnoses while providing provable and mathematically grounded privacy guarantees.
Hence, CADP pre-processed datasets may be used to increase anonymity of medical image data in the future. 
However, the level of required anonymity should be decided depending on the individual use case. 


Even for small $\epsilon < 1.0$ our method generates visually appealing results that can be used to train a classifier outperforming DP-SGD with the same privacy guarantees.
However, clipping of the latent space discards information for reconstruction.
In future work, it can be investigated how much information is lost to assure privacy.
Further, an in-depth exploration of the latent space can be conducted.

\section*{Acknowledgements} 
This research was supported by grants from the Klaus Tschira Foundation within the Informatics for Life framework, by the DZHK (German Centre for Cardiovascular Research), and by the BMBF (German Ministry of Education and Research).
The authors gratefully acknowledge the data storage service SDS@hd supported by the Ministry of Science, Research and the Arts  Baden-Württemberg (MWK) and the German Research Foundation (DFG) through grant INST 35/1314-1 FUGG and INST 35/1503-1 FUGG.




\bibliographystyle{splncs04}
\bibliography{bibliography}

\newpage
\appendix

\section{Network Architectures}
\label{sec:architectures}

\begin{table}[H]
    \centering
    \caption{Architectures of INN and classifier for different datasets. As optimizer Adam was used in all cases \cite{kingma2015adam}}\label{tab:architectures_inn}
    \begin{tabular}{L{2.5cm}|C{1.8cm}C{1.8cm}C{1.8cm}C{1.8cm}C{1.8cm}}
        Attribute &  MNIST & OCT & Chest X-ray & CelebA & Diabetes\\
        \toprule
        Coupling block & GIN & Glow & Glow & Glow & GIN \\
        \midrule
        Depth & $2 \times 4$ & $4 \times 4$ & $4 \times 4$ & $4 \times 6$ & -\\
        \midrule
        N blocks FC & 2 & 4 & 4 & 12 & 4 \\
        \midrule
        Input noise & 0.15 & 0.15 & 0.15 & 0.15 & 0.02\\
        \midrule
        Learning Rate & 5e-4 & 3e-4 & 3e-4 & 3e-4 & 1e-4 \\
        \midrule
        Batch Size & 512 & 64 & 64 & 16 & 442 \\
        \midrule
        Classifier depth & 2 & 5 & 5 & - & - \\
        \midrule
        C. Learning Rate & 5e-4 & 5e-4 & 5e-4 & - & - \\
        \midrule 
        C. Batch Size & 512 & 64 & 64 & - & -
    \end{tabular}
\end{table}





\section{Additional Results}
\vspace{-1cm}
\begin{figure}[H]
    \centering
    \includegraphics[width=0.85\textwidth]{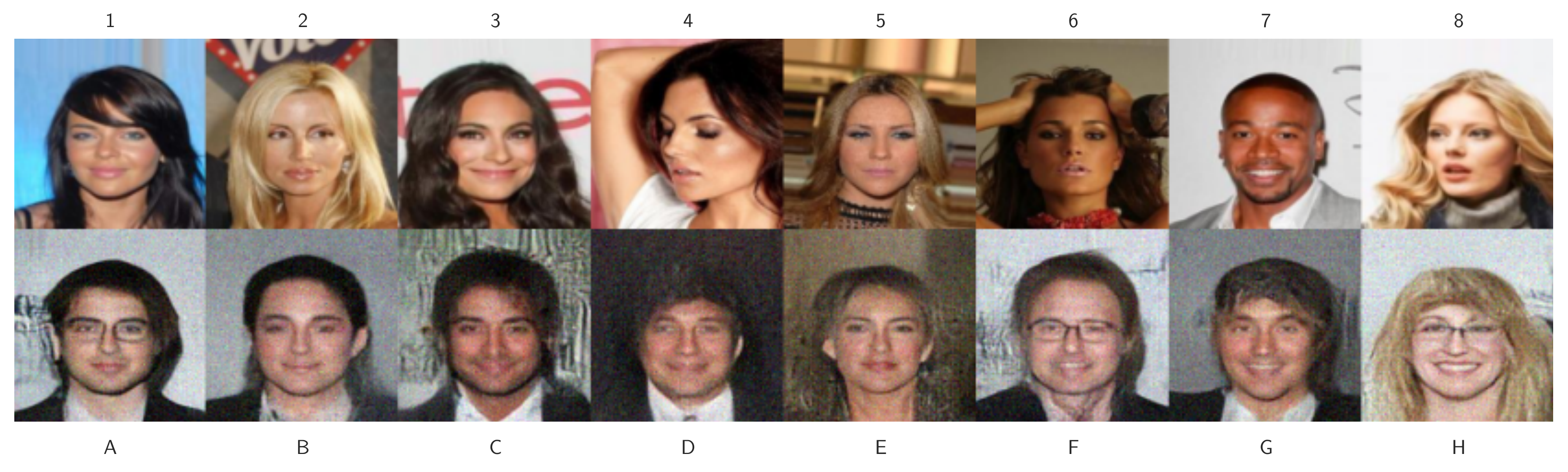}
    \caption{Can you guess who is who after applying CADP to faces \cite{liu2015celeba}? Solution: \rotatebox[origin=c]{180}{1 - H, 2 - D, 3 - F, 4 - A, 5 - E, 6 - B, 7 - G, 8 - C}
    }
    \label{fig:face_quiz}

    \vspace{.6cm}
    \centering
    \includegraphics[width=\textwidth]{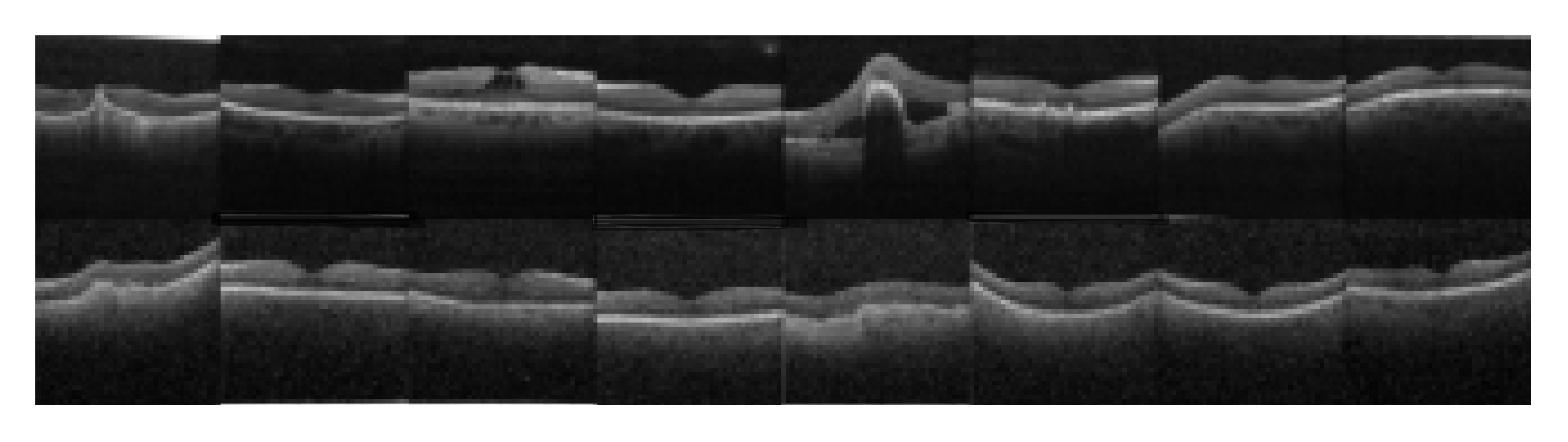}
    \caption{Additional results with content-aware differentially private images (bottom row) for random samples from OCT dataset (top row) with $\epsilon=1$ and sensitivity $s=0.5$ for all classes.
    An re-assignment of which ground truth image belongs to which reconstruction is difficult for a human observer.}
    \label{fig:oct_laplace_dp_sample}
\end{figure}

\begin{figure}[H]
    \centering
    \includegraphics[width=\textwidth]{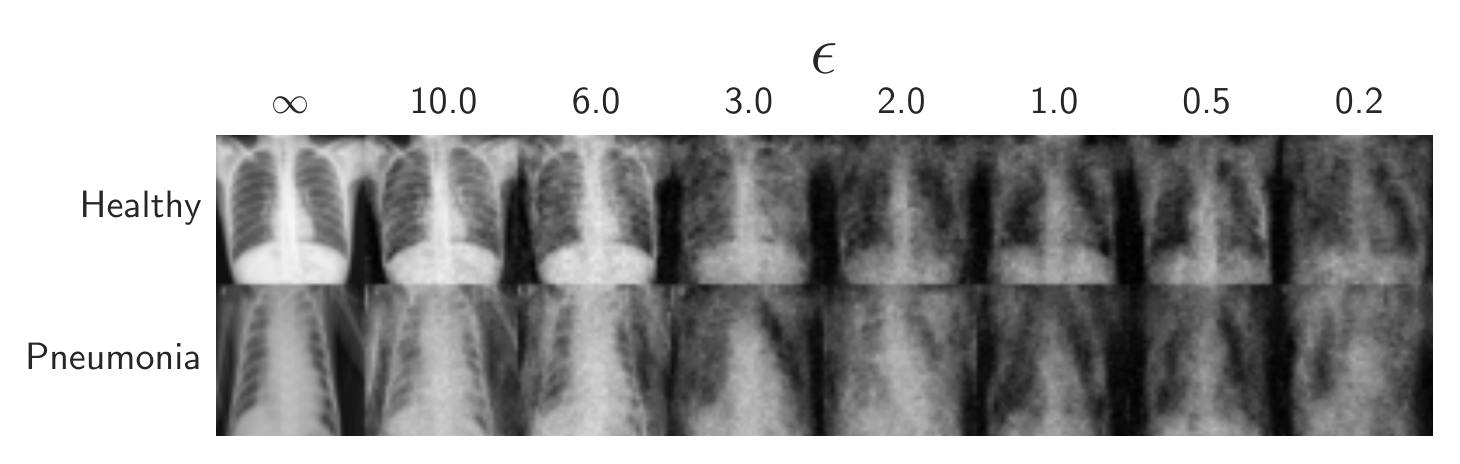}
    \caption{Content-aware differentially private images from Chest X-ray dataset with different $\epsilon$ for classes \textit{Pneumonia} and \textit{Healthy} \cite{kermanydata}. 
    The sensitivity is set to $\mathrm{min\,}(\epsilon/2, 4)$.
    In the case of chest x-ray images the reconstructions degrade with smaller $\epsilon$, which might be a consequence of the more complicated patterns.}
    \label{fig:xray_laplace_dp}
\end{figure}

\begin{figure}[H]
    \centering
    \includegraphics[width=\textwidth]{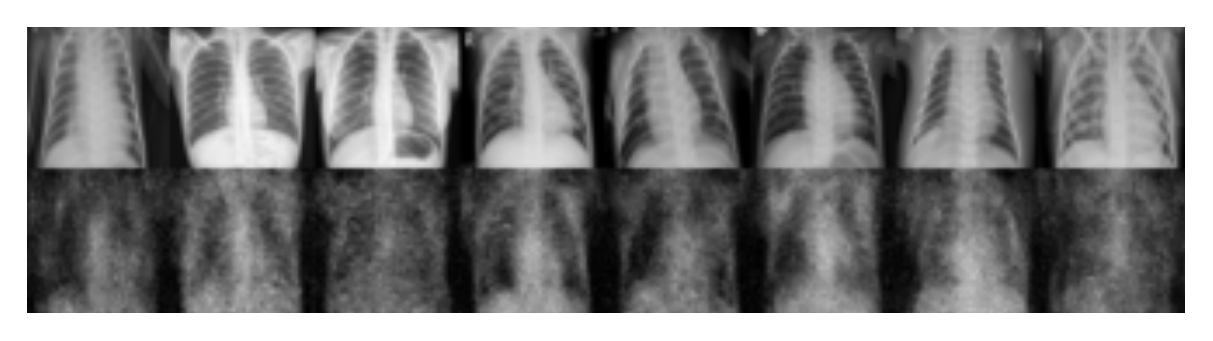}
    \caption{Additional results with content-aware differentially private images (bottom row) for random samples from chest x-ray dataset (top row) with $\epsilon=1$ and sensitivity $s=0.5$.
    Similar to Fig. \ref{fig:xray_laplace_dp} the images could be of higher quality for a human observer, but identification is impeded and classification results are on an acceptable level (Fig. \ref{fig:classifier})}
    \label{fig:xray_laplace_dp_sample}
\end{figure}

\begin{figure}[H]    
    \centering
    \includegraphics[width=\textwidth]{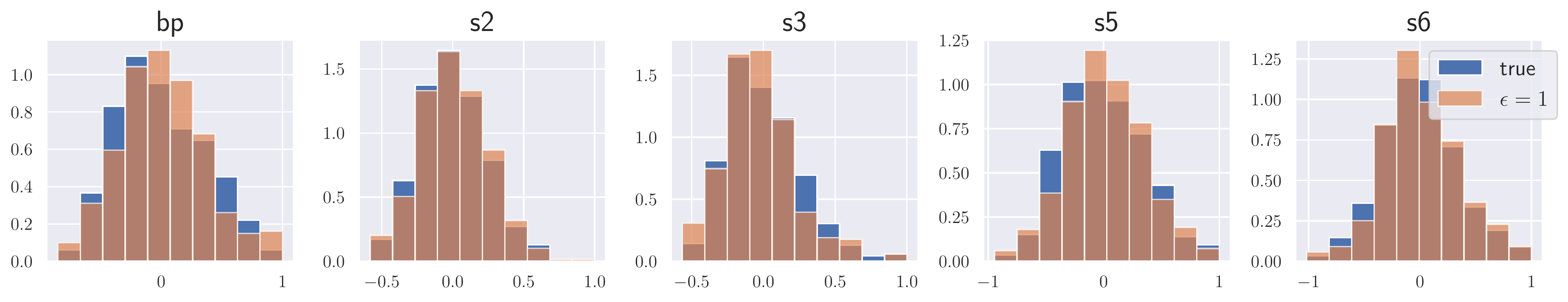}
    \caption{Additional features with content-aware differentially private data from diabetes dataset from \texttt{sklearn} with $\epsilon=1$ and sensitivity $s=1$ \cite{scikit-learn}.}
    \label{fig:categorical_dp_additional}
\end{figure}


\end{document}